\input{aipcheck}

\documentclass[
   ,final            
  ]
  {aipproc}

\layoutstyle{6x9}

\begin{document}

\title{Probability assignment in a quantum statistical model}


\author{L.F. Lemmens and Burhan Bakar}{
  address={ 
Universiteit Antwerpen\\
Departement Fysica\\
 Groenenborgerlaan 171 \\
B-2020 Antwerpen \\
Belgi\"e
}
}

\begin{abstract}
The evolution of a quantum system, appropriate to describe nano-magnets, can be mapped on a Markov process, continuous in $\beta$. The mapping implies  a probability assignment that can be  used to study the probability density (PDF) of the magnetization.  This procedure is not the common way  to assign probabilities, usually an assignment that is compatible with the von Neumann entropy is made.
Making these two assignments for the same system and comparing both PDFs, we see that they differ numerically. In other words the assignments lead to different PDFs for the same observable within the same model for the dynamics of the system. Using the maximum entropy principle we show that the assignment resulting from the mapping on the Markov process  makes less assumptions than the other one. 

Using a stochastic queue model that can be mapped on a quantum statistical model, we control both assignments on compatibility  with the Gibbs procedure for systems in thermal equilibrium and argue that the assignment resulting from the mapping on the Markov process satisfies the compatibility requirements.

\end{abstract}

\maketitle


\section{Introduction}

The statistical approach of a classical system relies on probability theory through the lack of knowledge on the initial conditions, leading to a well known  probability assignment within the Boltzmann-Gibbs framework.  In a quantum system the evolution is inherently stochastic and the knowledge on the initial condition of the system is also incomplete. Dealing with the statistical properties of a quantum system the question arises: where is the uncertainty in the formulation coming from? Does the generalization of the classical approach not give too much importance to the lack of knowledge on the initial conditions and do we incorporate correctly the propagation of uncertainty inherent to quantum theory?


 It is clear that we will not answer to these questions in general. We will only give  a few examples that illustrate the key-role played by the probability assignments. We start with a probability assignment that is not very well known. For  quantum models with a finite number or a countable number of states a mapping of the model and its evolution expressed by the parameter $\beta$ into a Markov chain with a continuous time dependence can be made. The mapping accomplishes  the probability assignment: from the  transition probabilities obtained by solving the Markov chain, the PDF of an observable at $\beta$ can be obtained given the initial PDF for the same observable using the Chapman-Kolmogorov equation. We will also consider the probability assignment that is compatible with the von Neumann entropy by taking the expectation value of an indicator for a specific state of the observable. For example assume that the observable $M$ can take the value $k$ and consider the indicator $\mathcal{I}(M=k)$. The PDF for a discrete observable $M$ is given by $\langle \mathcal{I}(M=k)\rangle$, with $\langle A \rangle= \textrm{Tr}A\exp(-\beta( H-F)) $ and $F$, the free energy of the model.


Once the probability assignments are made the Shannon entropy for both assignments can be calculated for the same system and compared. In the case of the mapping on the Markov chain the formalism allows a distinction between the uncertainty coming from the initial condition and the quantum evolution. In the common quantum statistical procedure such a separation of sources of uncertainty is less obvious. It should be noted that in both cases the Shannon entropy is calculated.

The paper is organized as follows, in the next section we give a short review of the main points in the mapping that lead to a Markovian representation of the quantum evolution of the systems. We sketch how it is done for the magnetic cluster and show that  the stochastic evolution of an ${\rm M/M/}\infty $-queue and the quantum evolution of a displaced harmonic oscillator can be regarded as representations of the same probabilistic system.
In the following section we compare the Shannon entropy calculated using the probability assignment based on the expectation value of an indicator  and compare this entropy with the Shannon entropy obtained by the Chapman-Kolmogorov equation for the Markovian representation of the system. 
In the last section a discussion on the mapping is given and some remarks on the relation between the blocking temperature of the magnetic system and the behavior of the entropy in terms of $\beta$ are given.

\section{The methodology}
In this section we indicate how a mapping on a Markov process helps us to calculate the PDF of a set of states $\mid k \rangle$, the eigenstates of an observable, order parameter etc, that undergo  an evolution  generated by an Hamiltonian. The cyclic property  of the trace allows you to write the partition function in any complete set of states, taking the states $\mid k \rangle$ the following expression is the expectation value of the indicator  $\mathcal{I}(M=k)$
\begin{equation}
p(k\mid\beta) =\frac{\langle k\mid \exp(-\beta H)\mid k \rangle}{Z(\beta)},
\label{PA}
\end{equation}
and is a probability assignment. The Shannon entropy calculated using (\ref{PA}) is sometimes called the information entropy \cite{PRL50-631, EuroLett67-700}.  In ref. \cite{PRE71-046109}, we compared the PDF of magnetization based on the assignment (\ref{PA}) with the PDF based on the Markov representation and we found that they are different. In order to indicate the possible origin of this difference we briefly  indicate the steps necessary to achieve the Markov representation. The equivalence in evolution between the quantum system and Markov process depends on a mapping of the evolution equations for the states $\mid k \rangle$ on a backward Kolmogorov equation. This allows to use the forward Kolmogorov equation to calculate the transition probabilities for evolution between the states. These transition probabilities denoted by $p(k \, \beta \mid k' \, 0)$ express the probability to observe the system in state $\mid k \rangle$ given $\beta$ when one knows it is in state $\mid k' \rangle$
initially. For any density of the initial states the final PDF can be obtained by conditioning:
\begin{equation}
p(k\mid \beta)=\sum_{k'}p(k \, \beta \mid k'  \, 0 )p(k' \mid 0).
\label{IC}
\end{equation}


Our objective  is to calculate the probability density of a complete orthonormal set of states $\left| k\right\rangle $ given some parameter $\beta$. In order to relate this probabilities for different values of the parameter we would like to know the conditional probability $p(k\,\beta\mid k' \beta')$ where $\beta >\beta'$. Let us emphasize that the conditional probability is a logical relation between the two states not a causal one. This allows to exchange the states and corresponding parameters using Bayes rule and obtain a description valid for  $\beta' >\beta$. 

We define  the following propagator of $H$, the Hamiltonian that generates the quantum evolution:\begin{equation}
K_{l\,k}\left( \beta \right) =\frac{g_{k}}{g_{l}}\left\langle k\right| {\exp }%
\left( -\beta H\right) \left| l\right\rangle ,  \label{PR}
\end{equation}
the states  $\left| k\right\rangle $  are the eigenstates of the operator $M$ not commuting with $H$.
The partition function $Z(\beta)$ needed to evaluate (\ref{PA}) can also be calculated from the diagonal
elements of (\ref{PR}). The propagator (\ref{PR}) satisfies the following
initial condition: $
\lim_{\beta  \downarrow 0}K_{l\,k}\left( \beta \right) = \delta_{l\,k}$. Using $
H\left| l\right\rangle =\sum\limits_{k\ne l}\left| k\right\rangle
\left\langle k\right| H\left| l\right\rangle +\left| 
l\right\rangle \left\langle l\right| H\left|
l\right\rangle$, the equation of motion of the propagator readily obtained: 
\begin{equation}
{\partial }_{\beta}K_{l\,k}\left( \beta \right)  =\sum\limits_{p\ne
l}q_{l\,p}K_{p\,k}\left( \beta\right) -v_{l}K_{l\,k}\left( \beta\right)   
-V_{l}K_{l\,k}\left( \beta \right) .
\label{Em}
\end{equation}
The following expressions have been introduced:
$q_{l\,p}$ given by $-\left\langle p\right| H\left| l\right\rangle \frac{g_{p}}{g_{l}}$ and $
v_{l}=\sum\limits_{p\ne l}q_{l\,p}$ while 
$ V_{l}$ is given by $\langle l \mid H \mid l \rangle -v_{l}$.
If the parameters $q_{l\,p}$ are positive  they can be considered as rates of a continuous time Markov chain, provided that $V_{l}$ is identically zero. In that case (\ref{Em}) is a backward Kolmogorov equation of a Markov chain \cite{BhatWay}. If $V_{l}$ is different from $0$ and positive, then the introduction of an additional state $\mid c\rangle$ can make (\ref{Em}) Markovian. This state is called the coffin state, it is a state characterized by  absorption and has the following properties by definition:
\begin{equation}
K_{c \, l}\left( \beta\right) = 0 , \quad K_{c\,c}\left( \beta\right) = 1 , \quad \sum_{k \neq c}K_{l\,k}\left( \beta\right)+ K_{l \, c}\left( \beta\right)=1.\label{Cons}
\end{equation}
Of course we have to check whether the conservation of probability in the presence of absorbing states (\ref{Cons}) can be satisfied for all values of $\beta$. The proof goes as follows: firstly we introduce the ground state $\langle \phi \mid$, secondly we shift the origin of spectrum of $H$ to $0$ and then we expand the ground state into the complete set $\langle k \mid$:
$
\langle \phi \mid=\sum_k g_k \langle k \mid
$.
From the conservation of probability follows:
$
g_l \propto \langle \phi \mid l \rangle
$,
and the proportionality factor has to be determined by the normalization.

The positivity of the rates   in the Markov process turns out to be the most demanding criterion for the existence of the mapping. We do not know the necessary   conditions, but it is easy to show that the conditions imposed by the Rokhsar-Kivelson mapping are sufficient \cite{RK-PRL61-2376, Cas-AnnPhys318-316}.


The first quantum model that we will study in a Markov representation is used to analyze the spectrum of some  nano-magnets. It is a finite state system described by the following Hamiltonian:
\begin{equation}
H=D({S}^{z})^2 +E(({S}^{x})^2-({S}^{y})^2)  +g\mu_{B}( {b_{z}}{S^{z}}+{b_{x}}{S^{x}}). \label{eq:GSH}
\end{equation}
This model belongs to the class of mean-field models and bears some resemblance with the Lipkin-Meshkov-Glick-model \cite{NPhys62-188}. When the Hamiltonian is rewritten in terms of spin lowering and raising operators, the generator $Q$ for the continuous time Markov process can be calculated using the techniques which are documented in ref. \cite{PRE71-046109}. 
Casting the Markov representation of ( \ref{eq:GSH})  into a matrix equation:
$
\partial_{\beta}\tilde{K}(\beta)=\tilde{Q}\tilde{K}(\beta), 
$ and
straightforward matrix methods can be used to obtain $K_{m\,m'}(\beta)$.
 In the spectral representation the propagator is given by:
\begin{equation}
\label{spr}
K_{m\, m'}(\beta)=\sum_{\nu} U_{m\,\nu}\exp(-\beta \epsilon_\nu)U^{-1}_{ \nu\,m'}
\end{equation}
where $ \epsilon_\nu $ is the $\nu$-th eigenvalue of the generator $\tilde{Q}$ and $ U_{l\,\nu}$ is an element of the modal matrix. The mapping and references to the physical origin of the model can be found in the paper: \cite{PRE71-046109}.

The second system is an ${\rm M/M/}\infty$ - queue. In the literature the symbol $X/Y/n/K/m/Z$ denotes a random service process in which customers join a queue if all servers are busy, with inter-arrival time distribution $X$, service time distribution $Y$, number of servers $n$, maximum number of customers allowed in the queue $K$, number of customers in the source $m$, servicing rule $Z$ \cite{BhatWay}. 
The backward Kolmogorov equation for this system, considering an evolution parametrised by $\beta$
is given by:
\begin{equation}
\partial_\beta K(n,\beta;n',0)=\gamma K(n+1,\beta;n',0)+\lambda nK(n-1,\beta;n',0)-(\gamma +\lambda n)K(n,\beta;n',0). \label{BKEMM}
\end{equation}
The parameter $\gamma$ is related to the rate that incoming costumers join the queue, while $\lambda$ is the rate to get a service for a customer that is in the queue and as a consequence leaves the queue.
Using the techniques explained in the preceding section it is straightforward to show that the equation (\ref{BKEMM}) is a mapping of a quantum system with the following ``Hamiltonian" :
\begin{equation}
H=\lambda A^{\dag}A-\gamma A-\lambda A{\dag}+\gamma,
\end{equation}
where the operators $A^{\dag}\, ,\, A$ are respectively raising and lowering operators. In the case that both rates take the same value, the quantum system turns out to be a displaced oscillator.
The methods to solve equation (\ref{BKEMM}) are algebraically equivalent to the methods used to introduce coherent states \cite{KlenSk}. Firstly a probability generating function is defined in the complex plane:
$
\Sigma_{n}(z, \beta)=\sum_{k}K(n,\beta ;k,0)z^k
$,
secondly the partial differential equation for the generating function is derived,
and solved using the method of characteristics.

Making use of an expansion in series the transition probabilities are obtained in terms of $g(\beta)= 1-\exp{(-\lambda(\beta))}$  by:
\begin{equation}
K(n,\beta ;m,0)=\exp{\left(-\frac{\gamma}{\lambda}g(\beta)\right)}\sum_{k}^{\min{(m,n)}}\pmatrix{n \cr k}\frac{(\frac{\gamma}{\lambda})^{m-k}}{(m-k)!}(1-g(\beta))^k\left(g(\beta)\right)^{n+m-2k}.
\end{equation}
Once the transition probabilities are known the propagation of the lack of knowledge on the initial conditions of the quantum system can be studied taking the stochastic nature of that evolution fully into account.

\section{Results and discussion}
We compare the Shannon entropy calculated on the basis of the two probability assignments introduced in the preceding section for the spin model and the  $ {\rm M/M/}\infty $ -- queue. 
\begin{equation}
\label{SHE}
S(\beta)=-\sum_m p(m\mid \beta)\ln p(m\mid \beta)
\end{equation}
We will indicate the quantities derived using the probability assignment based on the mapping into a Markov chain by an index $mc$ while those based on the expectation value of the indicator have an index $ie$. For he spin model we will consider two cases, firstly we consider a system consisting of a single large spin, typically used in connection with nano-magnets, and indicated by $LS$.  Secondly we will consider the many spin system, typically used to calculate the blocking temperature \cite{PRE71-046109} or other phase-transitions \cite{PRL49-478, PRA69-022107, PRA69-054101, PRL93-237204}, and indicated by $MS$.

For the $M/M/\infty$-queue we will use the maximum entropy principle to show that it is able to distinguish between different initial conditions. Furthermore we will calculate the Shannon entropy of the queue and of  the  "quantum`` model equivalent to the queue and show that the  entropies have completely different behavior with respect to $\beta$. In other words the predictions based on one of the assignments contradicts the predictions based on the other. 

At the end of the section we will draw some conclusions by  answering the question: what  probability assignment is compatible with the probability assignment for a quantum system in thermal equilibrium \cite{Feynman}, the so-called Gibbs assumption?
 
\subsection{The  spin model}
The single spin model describes the thermal evolution of a  large spin  with $N+1$ states. We assume that we have no prior knowledge about the preparation and chose a uniform density $p_{LS}(m\mid 0)= \frac{1}{N+1}$ as an initial condition. In this case equation (\ref{IC} ) is readily solved and
after normalization the entropy can be obtained by evaluating:
$p_{LS\,mc}(m\mid \beta)\propto \sum_k K(m,\beta;k)$.
Also the information entropy can be obtained using expression (\ref{SHE}) by replacing the expression  $p(m\mid \beta)$ by:$
p_{LS\,ie}(m\mid \beta)\propto K(m,\beta;m)
$.

\begin{figure}
\label{clusterfig}
  \includegraphics[height=.25\textheight]{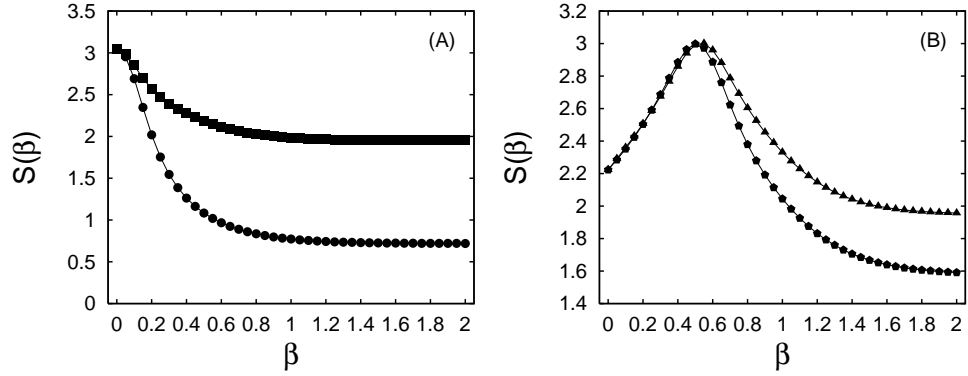}
  \caption{The entropy of the spin systems. 
(A): The thermal evolution of the entropy for two  probability assignments: the curve marked by black squares  incorporates the stochastic temporal evolution as well as the initial condition for the single spin model $S_{LS\,mc}(\beta)$, and is compared with the entropy $S_{LS\,ie}(\beta)$ marked by black dots. (B): The thermal evolution for two  probability assignments: the curve marked by black triangles  incorporates the stochastic temporal evolution as well as the initial condition for the many spin model $S_{MS\,mc}(\beta)$, and is compared with the  entropy $S_{MS\,ie}(\beta)$ of the same model marked by black pentagons.
The figure is made using $\frac{D}{k_B} $ as energy-unit and realistic values for $b$ and $E$, $b=E=.01\frac{D}{k_B}$.  }
\end{figure}

In figure (\ref{clusterfig}) we show the result for both probability assignments. From the comparison it is clear that the maximum entropy principle favors the probability assignment that incorporates the uncertainty about the initial condition and the stochastic evolution. 

We will compare now the entropy resulting from the two probability-assignments that we have used in the single spin model for a system that consists out of $N$ spin $\frac{1}{2}$ sites. The spins can flip up and down according to the  stochastic model generated by the same Hamiltonian. This model has the same spectrum as the single spin  system but the density of states is different. The multiplicity of a state with quantum number $m$ comes from the fact that there are ${N\choose m}$ realizations of this state. Due to the fact that the thermal evolution is independent of the site index the appropriate choice for the initial condition is a binomial density:
$
p_{MS}(m\mid 0)=\frac{1}{2^N}{N \choose m}
$.
Using this initial condition it is straightforward to obtain $p_{MS\,mc}(m\mid \beta)$ and to calculate the entropy (\ref{SHE}). 
The multiplicity of energy levels is taken into account  by incorporating it  in the probability assignment:
$
p_{MS\,ie}(m\mid \beta)\propto {N \choose m} K(m,\beta;m)
$, and adapting the normalization accordingly.

The obtained entropies are depicted in figure (\ref{clusterfig}), again it is clear that in the low temperature phase the maximum entropy principle \cite{Cat98} favors the solution of equation (\ref{IC}) and the appropriate initial condition. In the high temperature phase there is little difference between both probability assignments and the principle gives no compelling reason to prefer one assignment above the other. However only the Markov representation gives in both regimes an entropy that is either maximal or almost numerical equal to the maximum entropy.


\subsection{The queue}
According to the generating function of the moments  that interpolates for small $g(\beta) \approx 0$ from the generating function of a geometric density to that of a Poisson density at $g(\beta) \approx 1$, there are two possible candidates for the initial density of the queue i.e. the geometric density with a probability proportional to $p^n$ and a Poisson density $\exp(-\lambda)\frac{\lambda^n}{n!}$. In case of the geometric initial density we can use the generating function of the probability to obtain the solution of equation (\ref{IC}):
$
p_{geo}(n,\beta)\propto \Sigma_{n}(p, \beta)
$
while for the Poisson density we have to solve equation (\ref{IC}) numerically.

\begin{figure}
\label{CPGg}
  \includegraphics[height=.25\textheight]{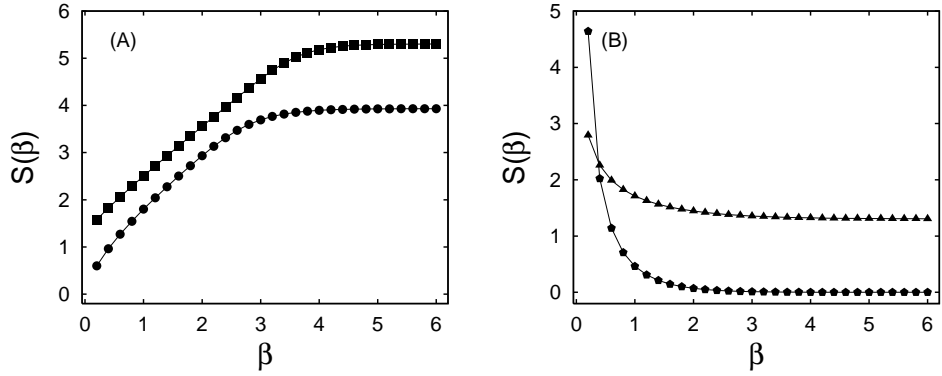}
  \caption{The entropies of an $M/M/\infty$-queue.
  In figure (A) the entropies resulting from a probability assignment that incorporates the stochastic evolution with two different initial conditions are compared. The entropy starting from the Poisson density (black squares)  is larger than the entropy starting from a geometric density (black dots). In figure (B) the entropies of an $M/M/\infty$-queue based on quantum statistical methods are shown. The Shannon entropy (black triangles) is obtained using the probability density obtained from diagonal elements of the propagator, the equilibrium entropy (black pentagons) is calculated directly from the partition function.}
\end{figure}

Once the probabilities are known  the expectation of their logarithm can be calculated to obtain their entropy. 
The results are shown in figure (\ref{CPGg}-A) and it is evident that the thermal evolution of the density with the Poisson density as initial condition leads to the maximum entropy. Indeed in the construction of the 
$M/M/\infty$-queue one assumes that the customers enter the facility according to a Poisson density. The initial geometric condition introduces a new assumption leading according to the maximum entropy principle to a lower entropy. 

Suppose that we insist on the use of the quantum statistical techniques \cite{JPA9-1469, PJPhy85, RMP70-979} to make statements about the entropy of the queue. In that case we will employ the resemblance with the displaced oscillator and in order to deal with a hermitian  model we chose the rates $\lambda$ and $\gamma$ equal. We can calculate an entropy using the partition function $S_{eq}=\beta(U-F)$  where $U$ is the internal energy and $F$ is the free energy. Using the diagonal part of the propagator:
$
p_{ie}(n,\beta) \propto K(n,\beta\mid n,0)
$
as a probability assignment an using expression (\ref{SHE}) is also a possibility. 

Both entropies are shown in figure (\ref{CPGg}-B). First we remark that the entropy calculated directly from the partition function differs from the  Shannon entropy, calculated from the diagonal parts of the propagator. Both show a thermal evolution that is in contradiction with the thermal evolution of the entropy of the queue. Clearly indicating that the probability assignment based on the stochastic evolution of the system is not necessarily compatible with the probability assignments often used in their quantum statistical counterparts. 
\subsection{What about Gibbs?}
The analysis of the entropies of the magnetic system could lead to the impression that one of the probability assignments leads to a reliable approximation of the other with respect to the thermal evolution of the system. The example with the queue excludes this impression. Invoking the maximum entropy principle is not adequate because the answer would depend on the precise value of $ \beta$ at high temperature the entropies shown in figure (\ref{CPGg}-B) are larger than those shown in figure (\ref{CPGg}-A) while for low temperatures it is the other way around. Therefore we have investigated what assignment is compatible with the one Gibbs used in the  case of thermal equilibrium. 

Let us consider the spectral representation of the propagator (\ref{spr}) and rewrite equation (\ref{IC}):
$
p(m,\beta)=\sum_{\nu}U(m,\nu)\exp(-\beta\epsilon_\nu)\sum_l U^{-1}(\nu,l)p(l,0).
$
Introducing 
$
w(\nu,\beta)=\sum_l U^{-1}(\nu,l)p(l,\beta)
$
we find that
\begin{equation}
w(\nu,\beta)=\exp(-\beta\epsilon_\nu)w(\nu,0).
\end{equation}
Noting that $\mid \nu \rangle$ is an eigenstate of the generator of the Markov process and therefore proportional to an eigenstate of the Hamiltonian, we see that equation (\ref{IC}) is compatible with the Gibbs assumption while the probability assignment based on the diagonal part of the propagator requires that the states $\mid k \rangle$ are eigenstates of $H$ in order to achieve compatibility.


\begin{theacknowledgments}
This work has been performed partly in the framework of the GOA BOF\ UA 2000 projects of the Universiteit Antwerpen.
\end{theacknowledgments}



\bibliographystyle{aipproc}   

\bibliography{057_lemmensbakar}

\IfFileExists{\jobname.bbl}{}
 {\typeout{}
  \typeout{******************************************}
  \typeout{** Please run "bibtex \jobname" to optain}
  \typeout{** the bibliography and then re-run LaTeX}
  \typeout{** twice to fix the references!}
  \typeout{******************************************}
  \typeout{}
 }

\end{document}